\pdfoutput=1

\documentclass{aa}

\usepackage{graphicx}
\usepackage[varg]{txfonts}
\usepackage{amsmath}
\usepackage{verbatim}
\usepackage{latexsym}
\usepackage{multirow}

\begin{document}

\title{The shrinking Sun: a systematic error in local correlation tracking of solar granulation}

\author{B. L\"optien\inst{1,2}
\and A.~C. Birch\inst{2}
\and T.~L. Duvall Jr.\inst{2}
\and L. Gizon\inst{2,1}
\and J. Schou\inst{2}}

\institute{Institut f\"ur Astrophysik, Georg-August Universit\"at G\"ottingen, 37077 G\"ottingen, Germany
\and Max-Planck-Institut f\"ur Sonnensystemforschung, Justus-von-Liebig-Weg 3, 37077 G\"ottingen, Germany}

\date{Received <date> /
Accepted <date>}

\abstract {Local correlation tracking of granulation (LCT) is an important method for measuring horizontal flows in the photosphere. This method exhibits a systematic error that looks like a flow converging towards disk center, also known as the shrinking-Sun effect.}
{We aim at studying the nature of the shrinking-Sun effect for continuum intensity data and at deriving a simple model that can explain its origin.}
{We derived LCT flow maps by running the local correlation tracking code FLCT on tracked and remapped continuum intensity maps provided by the {\it Helioseismic and Magnetic Imager} (HMI) onboard the {\it Solar Dynamics Observatory}. We also computed flow maps from synthetic continuum images generated from STAGGER code simulations of solar surface convection. We investigated the origin of the shrinking-Sun effect by generating an average granule from synthetic data from the simulations.}
{The LCT flow maps derived from HMI and from the simulations exhibit a shrinking-Sun effect of comparable magnitude. The origin of this effect is related to the apparent asymmetry of granulation originating from radiative transfer effects when observing with a viewing angle inclined from vertical. This causes, in combination with the expansion of the granules, an apparent motion towards disk center.}
{}
\keywords{Sun: granulation, sun: photosphere, radiative transfer, methods: data analysis}

\maketitle


\section{Introduction}
Local correlation tracking of granulation~\citep[LCT;][]{1988ApJ...333..427N} is an important method for measuring flows in the photosphere such as supergranulation or large-scale flows. Unfortunately, LCT suffers from a systematic error that looks like a flow converging towards disk center, also known as the shrinking-Sun effect~\citep{2004ESASP.559..556L}. This apparent flow is superimposed on the real solar flows measured by LCT and can reach up to one km/s. For some science objectives, this artifact can be removed by subtracting a temporal average from the LCT flow maps. However, this is not possible if the solar flows of interest are stationary in time. This is, for example, the case for the meridional flow.

Large-scale flows have already been measured with other feature tracking methods, such as supergranulation tracking~\citep[e.g.,][]{1980SoPh...66..213D,1990ApJ...351..309S,2012ApJ...760...84H}, tracking magnetic features~\citep[e.g.,][]{1996SoPh..163...21S} or tracking the motion of granules with the coherent structure tracking (CST) algorithm~\citep{2007A&A...471..687R,2012A&A...540A..88R,2013A&A...552A.113R}. A detailed comparison of the noise levels and systematics of the various methods is beyond the scope of this paper. Tracking granulation might produce different results for the large-scale flows than the other methods because it is more sensitive close to the surface of the Sun.

A good understanding of the shrinking-Sun effect is of particular importance for the upcoming {\it Solar Orbiter} mission~\citep{Yellowbook,Redbook,2015SSRv..196..251L}. Measuring the meridional flow at high latitudes using data provided by the {\it Polarimetric and Helioseismic Imager} (PHI) is one of the science objectives of the mission. If made with LCT, these measurements will be affected by the shrinking-Sun effect. Alternative methods that could be used for measuring the meridional flow, for example local helioseismology, suffer from systematic errors similar to the shrinking-Sun effect~\citep[see e.g.,][for an example for time-distance helioseismology]{2012ApJ...749L...5Z}. In case of {\it Solar Orbiter}, these errors will be affected by the varying image geometry - both the distance to the Sun and the viewing angle will change with time, leading to variations of the systematic errors.

The shrinking-Sun effect in local correlation tracking has so far only been studied by \citet{2004ESASP.559..556L} for the case of Dopplergrams obtained by the {\it Michelson Doppler Imager}~\citep[MDI;][]{1995SoPh..162..129S}. They suggest that the origin of this effect is related to the limited spatial resolution of MDI. They claim that LCT is more affected by granules moving towards disk center because these dominate the unresolved granulation pattern in the MDI data as they have a larger blueshift.

In this paper, we study the shrinking-Sun effect using continuum intensity data provided by the {\it Helioseismic and Magnetic Imager}~\citep[HMI;][]{2012SoPh..275..229S} and simulations of the solar surface convection~\citep[STAGGER code;][]{2009ASPC..416..421S,2009AIPC.1094..764S,2012LRSP....9....4S}. We compute LCT flow maps from both datasets using the local correlation tracking code FLCT~\citep{2004ApJ...610.1148W,2008ASPC..383..373F}. First, we study the dependence of the shrinking-Sun effect on the viewing angle and on spatial resolution. Then, we derive a simple model that explains the origin of the shrinking-Sun effect. It is based on the apparent asymmetry of granules originating from radiative transfer effects when observing with a viewing angle inclined from vertical. We also discuss various strategies for removing the shrinking-Sun effect.

\section{Data and methods}
\begin{figure*}
\centering
\includegraphics[width=17cm]{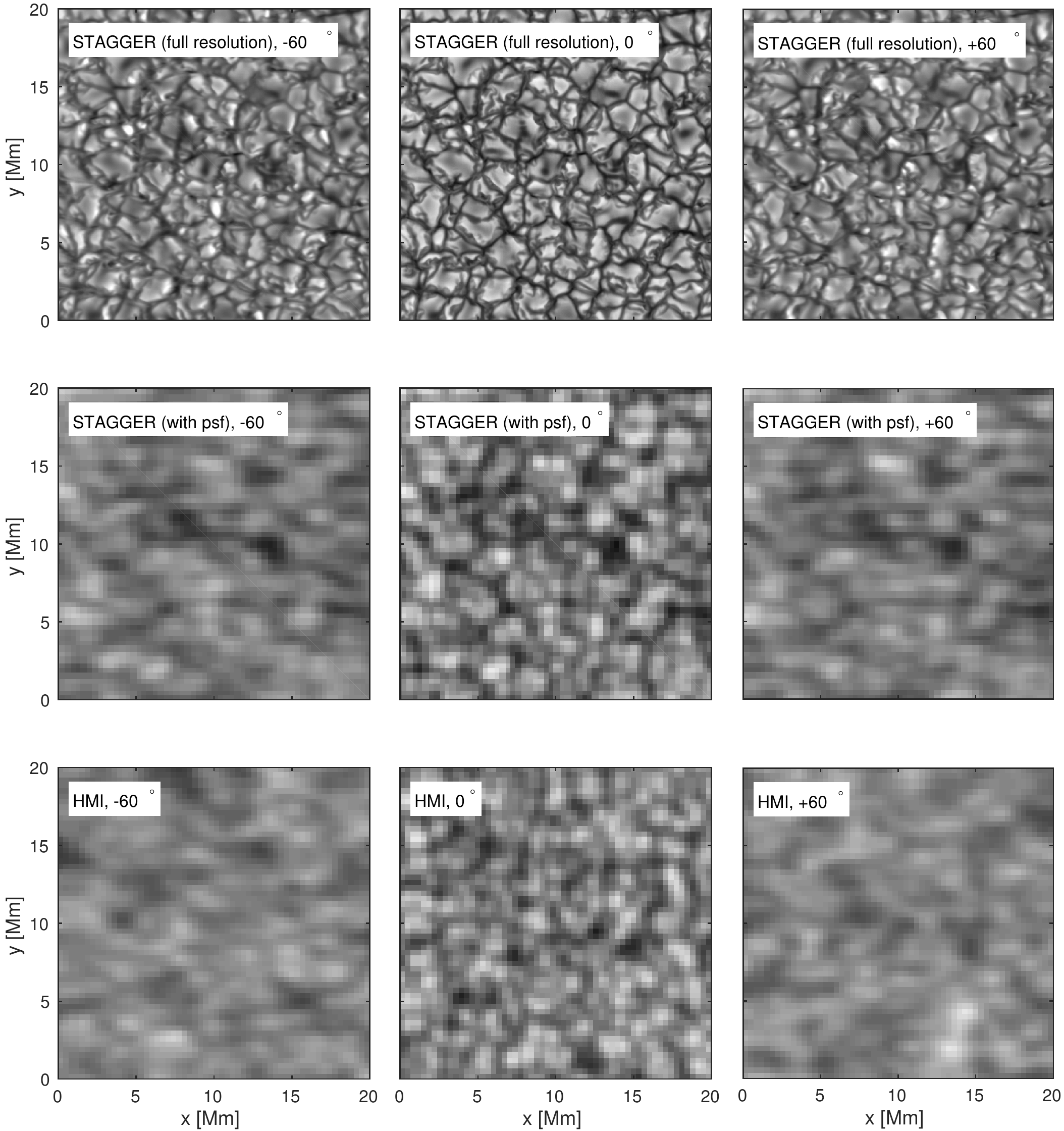}
\caption{Overview of the various datasets used in this study. {\it Top row:} continuum intensity maps derived from the STAGGER simulations (full spatial resolution), {\it middle row:} same as above but convolved with a point-spread-function~\citep[PSF; taken from][with $\gamma = 2.5$]{HMI_calibration}, {\it bottom row:} continuum intensity maps provided by HMI (different longitudes along the equator), obtained on 7 June 2010 ($B_0 = 0^\circ$). All datasets are shown for three different viewing angles, $-60^\circ$ ({\it left column}, right side points toward disk center), $0^\circ$ ({\it middle column}), and $+60^\circ$ ({\it right column}, left side points toward disk center). The HMI data and the synthetic data convolved with the PSF are plotted on the same scale.}
\label{fig:observables}
\end{figure*}

\begin{figure}
\resizebox{\hsize}{!}{\includegraphics{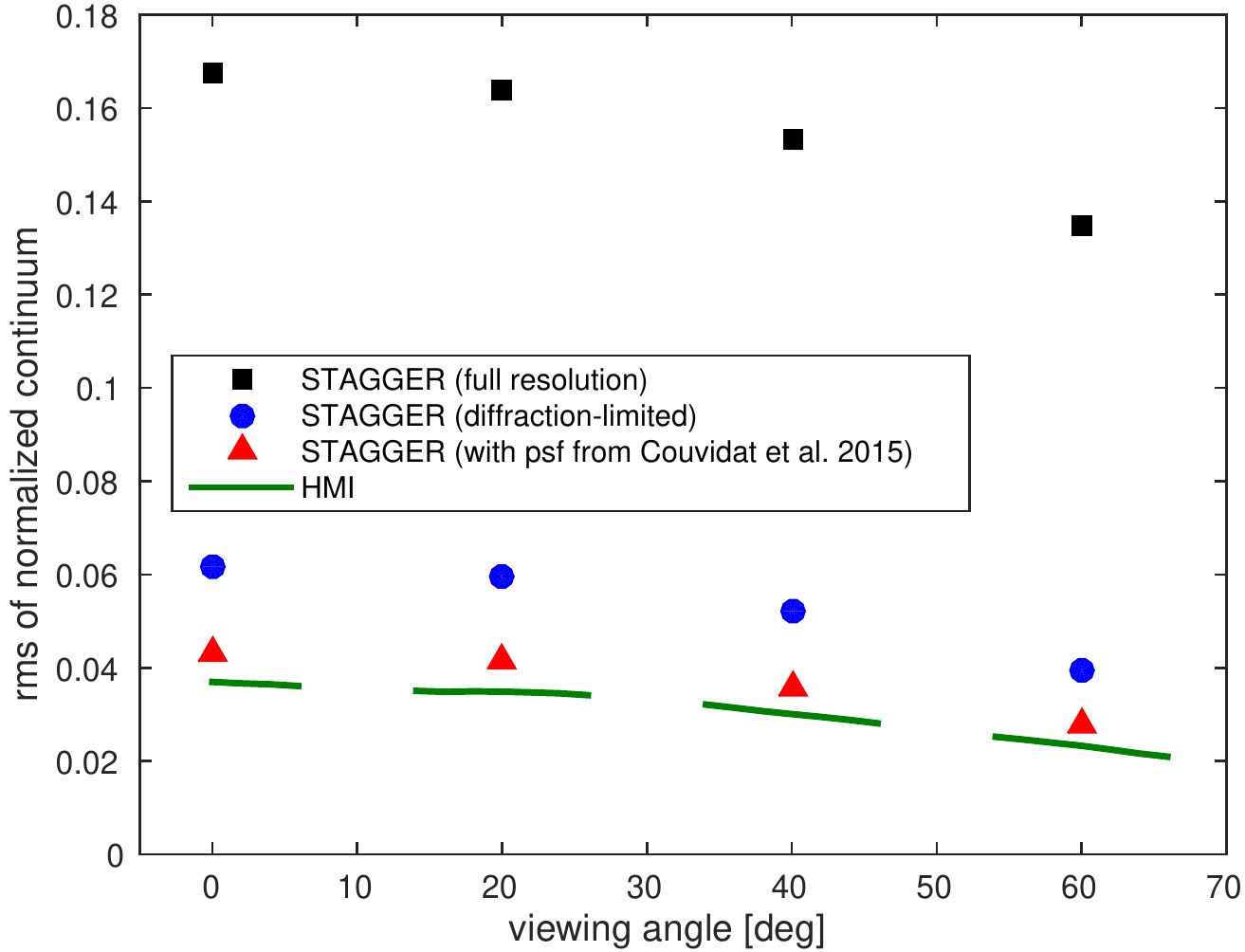}}
\caption{RMS of the normalized continuum intensity as a function of the viewing angle. We show results for four different datasets. {\it Black squares:} continuum intensity maps derived from the STAGGER simulations (full spatial resolution), {\it blue circles:} synthetic data corresponding to a diffraction-limited instrument, {\it red triangles:} STAGGER data convolved with the PSF from~\citet{HMI_calibration} with $\gamma = 2.5$, and {\it green line:} HMI data (along the equator, west of the central meridian). The error bars are smaller than the symbol size or the thickness of the line.}
\label{fig:RMS}
\end{figure}

\subsection{HMI data}
Our observational data consist of 74 days of tracked and remapped HMI continuum intensity images from 1 May to 13 July 2010 provided by \citet{2015A&A...581A..67L}. The data have a size of $178\times 178$~Mm ($512\times 512$~pixels, 348~km pixel size) and are centered around 7 latitudes along the central meridian ($-60^\circ$,$-40^\circ$,$-20^\circ$,$0^\circ$,$20^\circ$,$40^\circ$,$60^\circ$) and around 7 longitudes (same as for the latitudes) along the equator. The origin of the coordinate system is located at the intersection of the equator and the central meridian. The latitude increases northward, the longitude increases westward. The individual cubes have a cadence of 45~s, a length of 24~h, and are tracked at the Mt Wilson 1982/84 differential rotation rate~\citep{1984SoPh...94...13S} corresponding to the central latitude of each cube. In these data, the $B_0$-angle ranges between $-4^\circ$ and $+4^\circ$, with a mean value of zero.

The bottom row of Figure~\ref{fig:observables} shows examples of HMI continuum images for different longitudes along the equator.

\subsection{Synthetic continuum intensity images}
Our synthetic data are based on simulations of the solar surface convection with the STAGGER code~\citep{2009ASPC..416..421S,2009AIPC.1094..764S,2012LRSP....9....4S}. The simulation run used for this study has a size of \newline $96\times 96\times 20$~$\textmd{Mm}^3$, from which we use a time-series of snapshots of all physical parameters of the system with a cadence of 60~s and a duration of 6~hours. The horizontal resolution is 48~km and the vertical resolution varies between 12~km at the surface and 80~km at the bottom of the simulation box. The simulations exhibit a weak magnetic field corresponding to the quiet Sun. The average unsigned vertical field in the photosphere increases from 3~G at the beginning of the time-series to 4~G at the end of the time-series, the average unsigned horizontal field increases from 5~G to 7~G. The maximum field strength in the photosphere is 1.6~kG. The field is predominantly vertical and has no preferred horizontal direction (the azimuth of the horizontal field in the photosphere is uniformly distributed). The simulations also exhibit a mean flow in the $x$-direction that increases with height in the photosphere from about $-15$~m/s at $z = 0$ (defined as the average of the geometrical height where $\tau = 1$) to 0~m/s at $z = 400$~km.

We derive maps of the continuum intensity near the Fe~I~6173~\AA \ line (observed by HMI) from these simulations by computing the radiative transfer with the SPINOR code~\citep{2000A&A...358.1109F}. We generate continuum images for seven viewing angles $\varphi$ ($-60^\circ$,$-40^\circ$,$-20^\circ$,$0^\circ$,$20^\circ$,$40^\circ$,$60^\circ$). The viewing angle is constant throughout the entire simulation box. In order to reduce the computation time, we constrain ourselves to a $48\times 48$~$\textmd{Mm}^2$ subset of the simulations and use only pairs of consecutive snapshots of the simulations with the pairs separated by 30~min. We compute additional continuum intensity images for the snapshots three minutes after the first image of each pair (we do not use these images for LCT but for modeling the origin of the shrinking-Sun effect in Section~\ref{sect:cause}). Examples for the resulting continuum intensity images are shown in the top row of Figure~\ref{fig:observables}.

In the next step, we account for the influence of the HMI instrument. The main effect of the instrument is to decrease the spatial resolution. We evaluate the consequences of two different point-spread-functions (PSF). In the first case, we assume HMI to be diffraction-limited (the PSF is an Airy-function). The RMS of the resulting continuum intensity is significantly higher than for the actual HMI data (see Figure~\ref{fig:RMS}). Hence, we also test a more realistic PSF~\citep[equation $\lbrack 22 \rbrack$ with $\gamma  = 2.5$ in][]{HMI_calibration}, which is based on measurements on the ground. The data convolved with this PSF are in much better agreement with the HMI data, although the RMS is still a little bit too large (see Figure~\ref{fig:RMS}). This discrepancy could be caused, for example, by uncertainties in the ground-based PSF determination or by some degradation in space due to, e.g., darkening or temperature gradients across the front window. In case of HMI, the RMS also depends on disk position.
 For $\varphi = 60^\circ$, the RMS varies between $0.022$ (at the central meridian, north of the equator) and $0.025$ (at the central meridian, south of the equator). In all cases, the RMS of the HMI data is lower than the one of the synthetic data.

Finally, we account for the pixel size of HMI by convolving the data with a boxcar corresponding to the pizel size of the tracked and remapped HMI continuum images (348 km) and by subsampling the data. The middle row of Figure~\ref{fig:observables} shows examples for these continuum images.

\subsection{The FLCT code}
We perform local correlation tracking using the FLCT code~\citep[Fourier Local Correlation Tracking;][]{2004ApJ...610.1148W,2008ASPC..383..373F}. FLCT provides 2D maps of horizontal flows on the solar surface by tracking the motion of the solar granulation pattern.

For each pair of consecutive intensity images, the code estimates the horizontal flows by computing the cross-correlation for individual subimages. The size of these subimages is given by the parameter $\sigma$. It determines which features are being tracked and sets the spatial resolution of the resulting flow maps. Here we use $\sigma \approx 2.1$~Mm (6~pixels for the HMI and the synthetic HMI data, 44~pixels for the STAGGER data with full spatial resolution). The outputs of the code are flow maps both for flows in the north-south direction ($v_y$, pointing northward) and the east-west direction ($v_x$, pointing westward).

In case of the synthetic data, we run the FLCT code on the individual pairs of continuum images that we have computed with SPINOR. The images within the pairs are separated by 60~s, the individual pairs are separated by 30~min. This results in one LCT flow map for every 30~min of the simulations. This cadence is sufficient for this study because the shrinking-Sun effect causes a very strong signal in the LCT flow velocities. In case of the HMI data, we make use of the full cadence of the data (45~s).

LCT, in general, underestimates the flow speeds on the Sun~\citep[see e.g.,][]{2007SoPh..241...27S,2013A&A...555A.136V}. We correct for this by generating calibration data~\citep[see][]{LCT_paper}. These consist of continuum intensity images to which we add a constant flow by shifting the individual images in one direction using Fourier interpolation. The factors by which LCT underestimates the real flows range from $0.65$ in case of HMI data $67^\circ$ away from disk center to $0.98$ in case of the synthetic data with full spatial resolution for an observation at disk center.

\section{Observing the shrinking-Sun effect}
\begin{figure*}
\centering
\includegraphics[width=17cm]{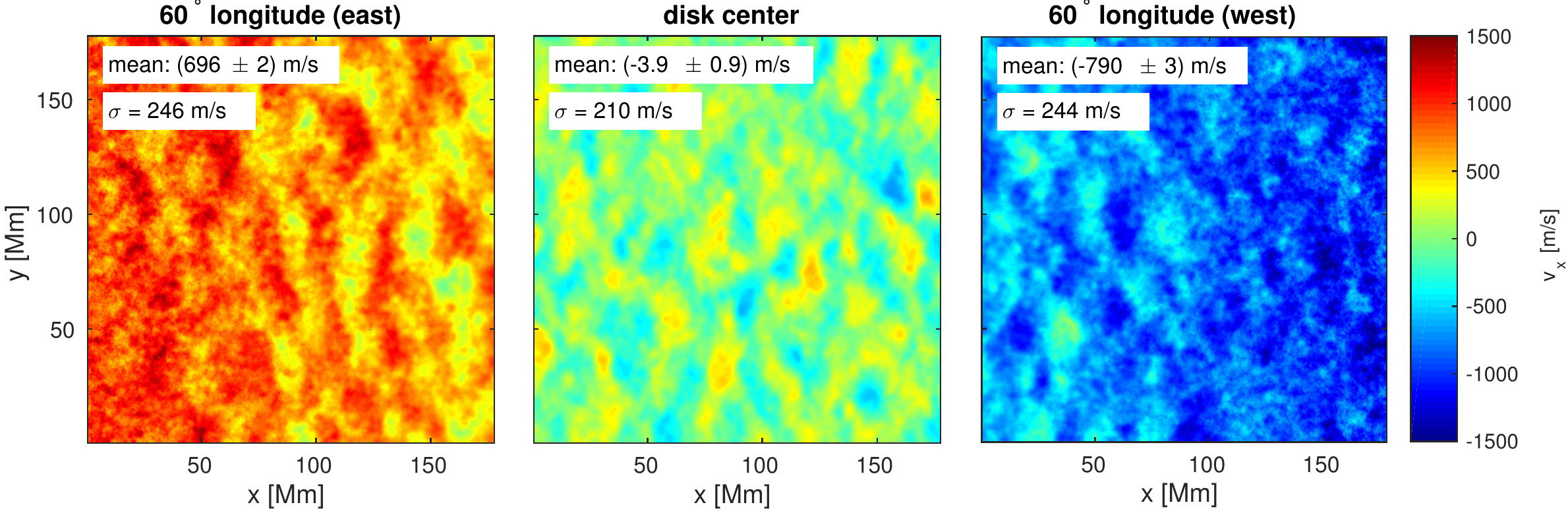}
\caption{LCT flow maps derived from 24~h time-series of HMI continuum intensity images obtained on 7 June 2010 ($B_0=0^\circ$) for different longitudes along the equator. {\it left:} $60^\circ$ east of the central meridian, {\it center:} disk center, and {\it right:} $60^\circ$ west of the central meridian. We show the velocity directed in the east-west direction ($v_x$, pointing westward). The numbers in each panel give the mean and the standard deviation of the velocities. All datasets clearly show supergranulation, however, when observing away from disk center, an apparent large-scale flow is superimposed on the flow maps, reaching up to one~km/s. This is the shrinking-Sun effect.}
\label{fig:flow_maps}
\end{figure*}
Figure~\ref{fig:flow_maps} shows examples for LCT flow maps derived from HMI continuum data. All of them clearly show supergranulation flows. However, a large-scale flow that looks like a converging flow towards disk center is superimposed on top of the supergranulation. This ``flow'' is the shrinking-Sun effect. Its magnitude increases with distance from disk center, reaching up to one~km/s. The shrinking-Sun effect is also present in the flow maps derived from the synthetic continuum images, both for the data with and without the PSF. 

\begin{figure}
\resizebox{\hsize}{!}{\includegraphics{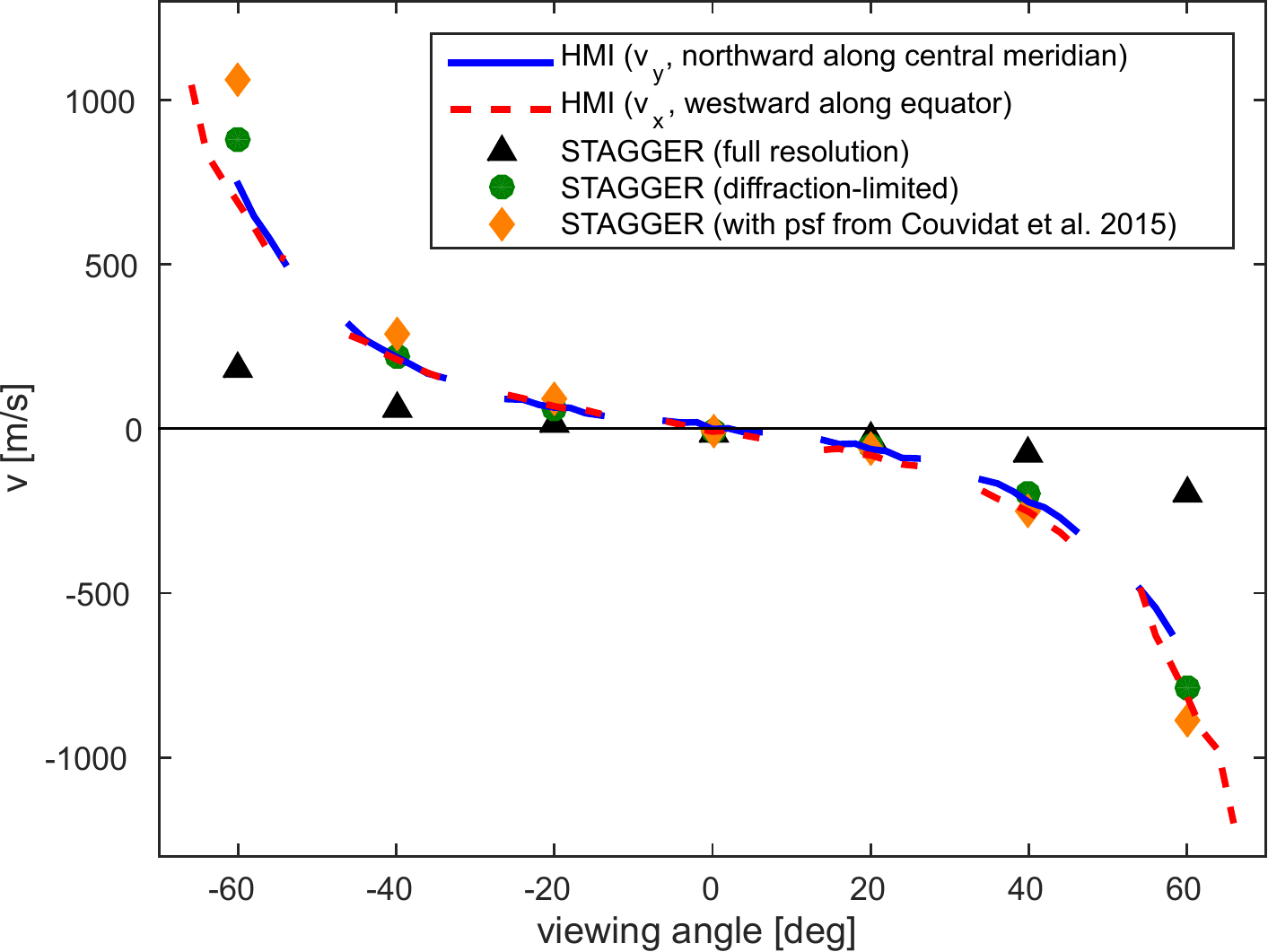}}
\caption{Flow velocities along the central meridian and along the equator as a function of the distance to disk center for different datasets. {\it Solid blue line:} HMI data along the central meridian, {\it dashed red line:} HMI data along the equator, {\it black triangles:} STAGGER data with full spatial resolution, {\it green circles:} STAGGER data convolved with a PSF (Airy-function), and {\it orange diamonds:} STAGGER data convolved with the PSF from~\citet{HMI_calibration}. The HMI data are averaged over latitude/longitude and time, the STAGGER data are averaged over $x$, $y$, and time. The error bars are smaller than the symbol size or the thickness of the line.}
\label{fig:shrinking_sun}
\end{figure}

Figure~\ref{fig:shrinking_sun} shows the flow velocities along the central meridian and along the equator as a function of the distance to disk center, after averaging over time and latitude/longitude. Since the mean value of the $B_0$-angle of our data is zero, the intersection of the equator and the central meridian (the origin of our coordinate system) is located at disk center. Both the latitude and the longitude of the averaged dataset give the distance to disk center. This allows to directly compare the shrinking-Sun effect along the central meridian and along the equator. We correct the data along the central meridian for variations of the $B_0$-angle (between $-4^\circ$ and $+4^\circ$ in our data). The shrinking-Sun effect depends on the distance to disk center but our intensity cubes are tracked at fixed latitudes. For each day of the time-series, we determine the distance of our intensity datacubes to disk center by subtracting the $B_0$-angle to the latitude of the cube. Then, we interpolate 
the LCT velocities from a grid in latitude to a grid in distance to disk center using a linear interpolation.

The velocities shown in Figure~\ref{fig:shrinking_sun} are a superposition of large-scale flows on the Sun (the residual signal from differential rotation after the tracking for the data along the equator, and the meridional flow for the data along the 
central meridian) and the shrinking-Sun effect. We also show the averaged flow velocities derived from the synthetic continuum intensity images. When the continuum images are convolved with a PSF, the velocities derived from the synthetic data are in good agreement with the observations, suggesting that the STAGGER simulations are suitable for studying the origin of the shrinking-Sun effect. There are, however, large differences between the two PSFs tested in this study. The PSF from~\citet{HMI_calibration} leads to a significantly stronger shrinking-Sun effect than expected for a diffraction-limited instrument ($\sim 100$~m/s difference at $60^\circ$). This indicates that the shrinking-Sun effect is extremely sensitive to changes of the spatial resolution.

The shrinking-Sun effect is nearly antisymmetric across the central meridian but exhibits an asymmetry across the equator. West of the central meridian, the shrinking-Sun effect is stronger than east of the central meridian (687~m/s at $60^\circ$ east of the central meridian and -814~m/s at $60^\circ$ west of the central meridian, see also Figure~\ref{fig:flow_maps}). This asymmetry increases with increasing distance to disk center. Similarly, for the synthetic data derived from the STAGGER simulations, the shrinking-Sun effect is stronger for negative viewing angles than for positive ones. When using the PSF from~\citet{HMI_calibration}, the mean velocity for $\varphi=-60^\circ$ is $(1060\pm 68)$~m/s but for $\varphi = +60^\circ$, it is only $(-890\pm 41)$~m/s. The reasons for these asymmetries are currently unclear.

\section{Modeling the shrinking-Sun effect}\label{sect:cause}
\subsection{The average granule}
\begin{figure*}
\centering
\includegraphics[width=17cm]{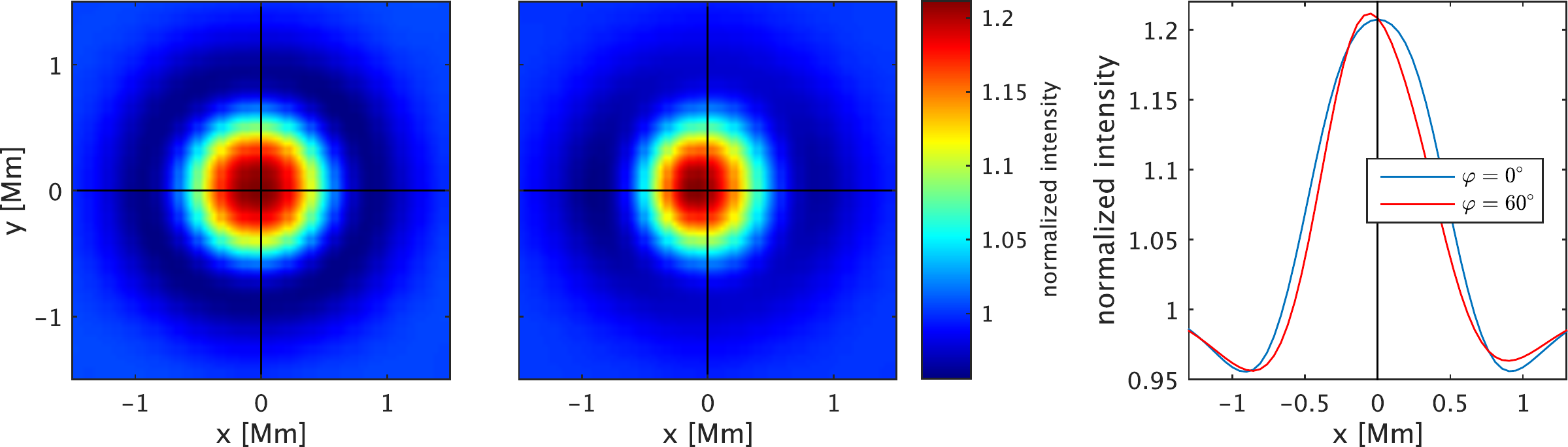}
\caption{Average intensity for granules derived from the STAGGER simulations (full spatial resolution). {\it Left:} average granule at disk center, {\it center:} average granule for $\varphi=60^\circ$, and {\it right:} cuts at $y=0$ for the average granule at disk center ({\it blue curve}) and $\varphi=60^\circ$ ({\it red curve}). The positive $x$-axis points away from disk center. In the plot on the {\it right}, the error bars are smaller than the line width.}
\label{fig:averaged_granule}
\end{figure*}

The shrinking-Sun effect is present in the synthetic data with full spatial resolution, so its origin cannot be related solely to instrumental effects (e.g., limited spatial resolution). In these data, it causes an almost constant offset in the velocity for flow maps for an observation away from disk center relative to the flow maps obtained at disk center. Apart from this offset, there are almost no differences between the flow maps derived from the synthetic data for different viewing angles.

Increasing the viewing angle increases the formation height of the continuum. The formation height is different for the side of the granules that points towards disk center and the side pointing toward the limb. The light emerging from the side of the granule that points towards disk center crosses the neighboring intergranular lane, where the opacity is low. Hence, radiation from deeper layers in the granules, where the temperature is higher, can emerge from the photosphere. This leads to higher continuum intensities. At the other side of the granule, the continuum forms higher in the atmosphere, at lower temperatures. So, the side of granules that points towards disk center is brighter than the one directed towards the limb. This effect is clearly visible in Figure~\ref{fig:observables}. As this is the only asymmetry between positive and negative viewing angles in the STAGGER data with full spatial resolution, it must be responsible for the shrinking-Sun effect.

Studying the connection between the shrinking-Sun effect and the apparent asymmetry of granules away from disk center is difficult due to the large variety of granular shapes. We circumvent this issue by constructing average granules from the STAGGER simulations and the HMI data for each viewing angle. We identify the centers of the granules as the local maxima of the intensity (in case of the synthetic data with full spatial resolution, we first convolve the  continuum intensity maps with a Gaussian with a width $\sigma_G = 9$~pixels, corresponding to 428~km, in order to smooth the structure of the granules). Then, we add up the intensities of the individual granules identified in this manner (in case of the synthetic data with full spatial resolution, we use the undegraded data). Figure~\ref{fig:averaged_granule} shows the resulting average granules at disk center and $\varphi = 60^\circ$ derived from the synthetic data with full spatial resolution. As expected, for $\varphi = 60^\circ$, the 
side of the granule pointing towards disk center is brighter than the opposite side. The average granule at disk center is azimuthally symmetric.

\subsection{Time-evolution of the average granule}
\begin{figure*}
\includegraphics[width=17cm]{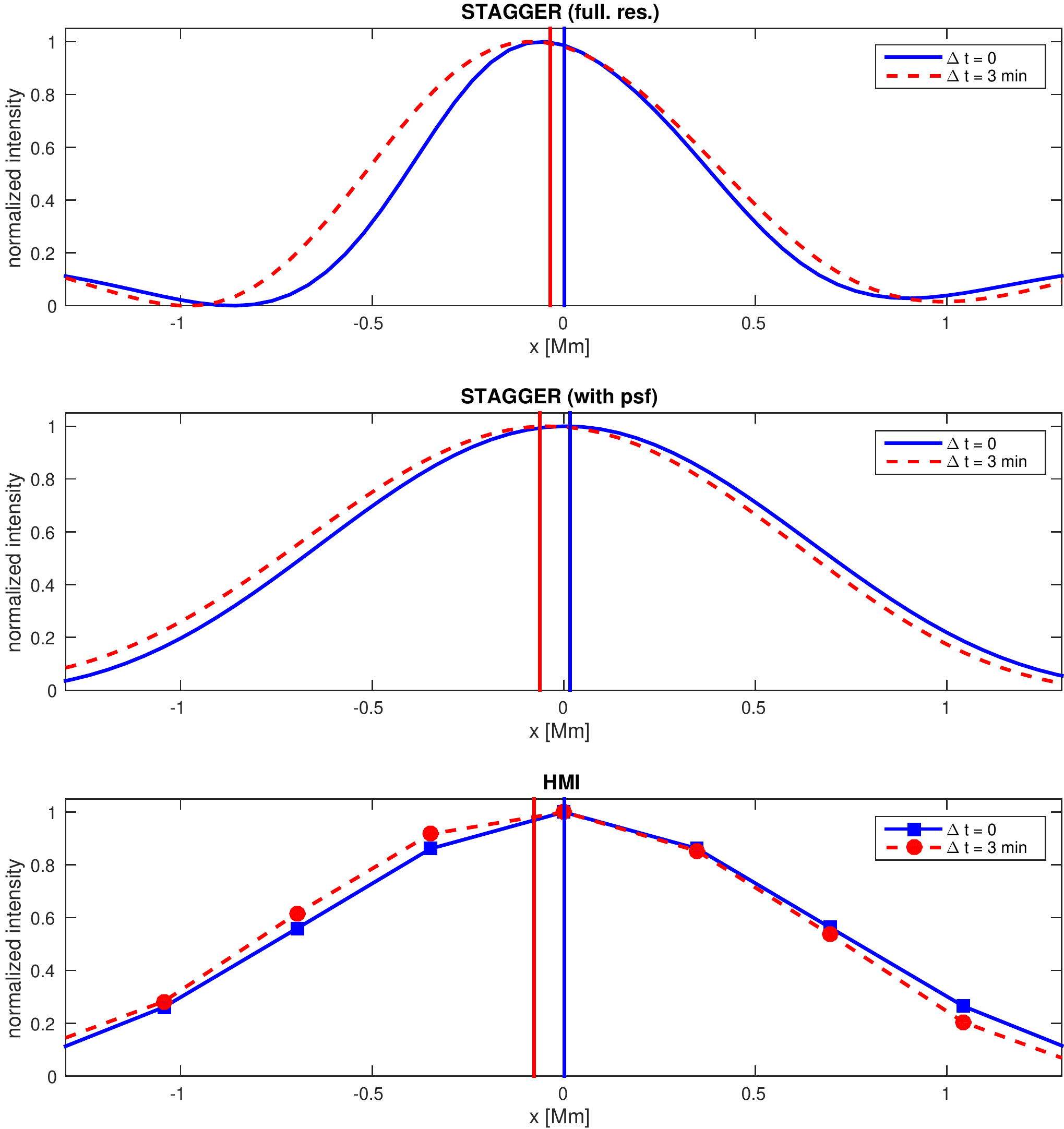}
\caption{Time-evolution of the average granule for a viewing angle of $60^\circ$ both for synthetic data and HMI data. We identify the positions of the granules in one image ($\Delta t = 0$) and then compute the average granule from this image and another one three minutes later, using the same positions of the granules. {\it Top row:} STAGGER continuum data with full spatial resolution, {\it middle row:} STAGGER data convolved with the PSF from~\citet{HMI_calibration}, {\it bottom row:} HMI continuum data (along the equator). {\it solid blue line:} average granule for $\Delta t = 0$, {\it dashed red line:} average granule for $\Delta t = 3$~min. The vertical lines show the center-of-gravity of the average granule for $\Delta t = 0$ ({\it black line}) and $\Delta t = 3$~min ({\it red line}). The $x$-axis points away from disk center. The intensity is normalized to values between 0 and 1 for all granules shown here. The error bars are smaller than the symbol size or the thickness of the lines.}
\label{fig:averaged_granule_time}
\end{figure*}

Investigating the origin of the shrinking-Sun effect requires not only an understanding of the apparent asymmetry of granulation for high viewing angles but also a model of the evolution of granulation. Granules move due to their proper motion and many granules also expand during their lifetime. This is important for LCT because LCT does not work on individual intensity images but on a time-series of intensity images. However, since LCT tracks the motion of the granulation pattern, it is not sensitive to the evolution of individual granules but only to the evolution of the granulation pattern as a whole. This consists of many individual granules at various evolutionary stages. Only the overall evolution of these individual granules is important for LCT, allowing us to study the influence of the evolution of granules on LCT using our simple model of the average granule.

We study the time-evolution of the average granule by first determining the positions of the individual granules in one image. Then, we derive the average granule from this image and the subsequent ones, using the same positions of the granules. Figure~\ref{fig:averaged_granule_time} shows the time-evolution of the average granule both for the synthetic data~\citep[with full spatial resolution and with the PSF from][]{HMI_calibration} and for the HMI data. At disk center (not shown here), the average granule expands with time. The expansion is azimuthally symmetric, since there is no preferred direction. However, away from disk center, the evolution is not azimuthally symmetric. At $\varphi = 60^\circ$, the granules still expand but also appear to be moving towards disk center. At $\varphi = -60^\circ$ (not shown in the figure), this apparent motion is directed in the opposite direction. This antisymmetry between positive and negative viewing angles shows that this motion is an artifact 
caused by the inclined viewing angle (the simulations also do not exhibit a strong large-scale flow). The reason for this shift is related to the asymmetry of granulation when observing close to the limb: the granules are brighter on the side pointing towards disk center. The bright side of the granules contributes more to the intensity of the average granule and due to the expansion of the granules, this side moves towards disk center. The combination of the asymmetry in visibility and the expansion causes the apparent motion of the average granule towards disk center.

\begin{figure}
\resizebox{\hsize}{!}{\includegraphics{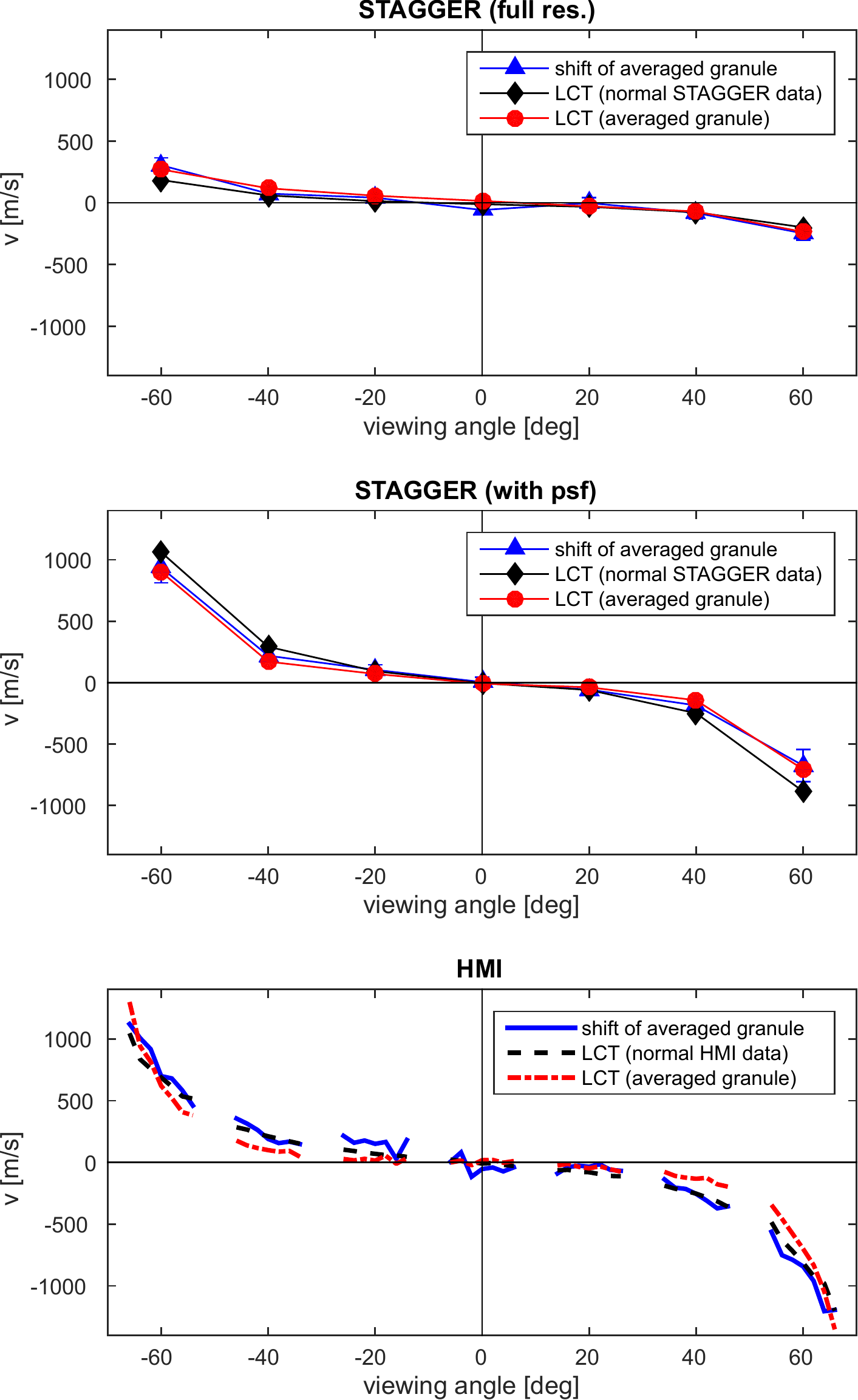}}
\caption{Comparison of the apparent motion of the average granule and the shrinking-Sun effect. {\it Top:} synthetic data with full spatial resolution, {\it center:} synthetic data convolved with the PSF from~\citet{HMI_calibration}, and {\it bottom:} HMI data (along the equator). We compare the motion of the average granule ({\it blue triangles}) with the LCT velocities of the nominal continuum images ({\it black diamonds}), and the LCT velocities when running the FLCT code directly on the average granule ({\it red circles}). When no error bars are shown, they are smaller than the symbol size.}
\label{fig:compare}
\end{figure}

This apparent motion is directly connected with the shrinking-Sun effect. Figure~\ref{fig:compare} compares the velocity of this apparent motion (defined as the motion of the center-of-gravity of the average granule between two consecutive intensity images) with the magnitude of the shrinking-Sun effect. We measure the magnitude of the shrinking-Sun effect in two different ways: we run the FLCT code both on the nominal intensity images but also directly on the average granule. The apparent motion of the average granule is in good agreement with the shrinking-Sun effect. The correlation coefficient between the shift of the average granule and the LCT velocities derived from the nominal intensity images is $0.982$ for the synthetic data with full spatial resolution, $0.998$ for the degraded data, and $0.992$ for the HMI data. This suggests that the shrinking-Sun effect can be explained by the shift of the average granule due to the asymmetry of the granules induced by radiative transfer effects and their 
evolution.

\section{Discussion and conclusion}
This paper explains the origin of the shrinking-Sun effect as a combination of the apparent asymmetry of granules when observing away from disk center and the evolution of granules. The cross-covariances computed by the LCT algorithm are more affected by the bright sides of the granules than by the darker sides. The motion of these bright sides is a superposition of the motion of the entire granule and motions originating from the evolution of the granules. Many granules, especially large ones, expand during their lifetime~\citep{1999ApJ...515..441H}, pushing their bright sides towards disk center. This causes an offset in the derived LCT flow velocities towards disk center as observed in the STAGGER simulations. Expanding granules have expansion velocities of about one km/s~\citep{2012A&A...537A..21P}. This is in good agreement with the maximum velocities caused by the shrinking-Sun effect.

Our results are a first step towards understanding the shrinking-Sun effect. They do not point toward a method for removing this effect from the flow maps, as required for measuring the meridional flow. This is challenging because the velocities caused by the shrinking-Sun effect are two orders of magnitude higher than the meridional flow. In addition, the shrinking-Sun effect is sensitive to changes of the spatial resolution. Even the difference between an Airy-function and the PSF from~\citet{HMI_calibration} has a large influence on the shrinking-Sun effect (see Figure~\ref{fig:shrinking_sun}). Variations of the PSF with time or over the field-of-view will severely affect any potential correction of the shrinking-Sun effect. This will be a problem for {\it Solar Orbiter}, since the image geometry will change during the mission. The distance of the spacecraft to the Sun will vary with time, which will affect the spatial resolution of the obtained images. The shrinking-Sun effect might also depend on other quantities that affect granulation, e.g., the magnetic field.

In principle, there are at least a couple of methods for subtracting the shrinking-Sun effect from the data. One possibility would be to derive a model of the shrinking-Sun effect. In this paper, we used synthetic data derived from simulations of solar surface convection to model the shrinking-Sun effect. Although this model is in good agreement with the observations, it does not yet have the accuracy required for measuring the meridional flow. These differences could be caused, for example, by an inaccurate PSF. Another way of modeling the shrinking-Sun effect would be to predict it from some properties of the granulation pattern, similar to the motion of the average granule shown in Figure~\ref{fig:compare}. Unfortunately, the motion of the average granule cannot be used for correcting for the shrinking-Sun effect because it is also sensitive to the large-scale flows.

Another approach would be to make use of the different symmetries of the shrinking-Sun effect and the large-scale flows. The large-scale flows depend only on latitude, the shrinking-Sun effect is mostly a function of the distance to disk center. Similar symmetries arise in measurements using Doppler images, where the convective blueshift is superimposed on the large-scale flows. The different symmetries of the blueshift and the large-scale flows can be used to remove this artifact \citep[e.g.,][]{1979SoPh...63....3D,1984SoPh...94...13S,1992SoPh..137...15H,2015arXiv151106500S}. In local helioseismology, a similar approach is used to remove a systematic bias from measurements of the meridional flow~\citep{2012ApJ...749L...5Z}.

All these methods assume the bias to be removed to depend only on the distance to disk center. Unfortunately, the shrinking-Sun effect exhibits an asymmetry around the central meridian, which is not understood yet. Due to this asymmetry, the shrinking-Sun effect depends not only on the distance to disk center but also on the azimuth. Hence, the origin of this asymmetry needs to be understood before removing the shrinking-Sun effect using symmetry properties.

The origin of the asymmetry of the shrinking-Sun effect is unclear. The shrinking-Sun effect is extremely sensitive to variations of the PSF. If the PSF of HMI does not only depend on the distance to disk center, it could cause the asymmetry. This would also explain the observed dependency of the granulation contrast on azimuth. There is only an asymmetry in the east-west direction but not in the north-south direction. This points towards a connection with solar rotation. Since the asymmetry increases with increasing distance to disk center, it cannot be caused by an inaccurate tracking rate. One possible way how rotation could be responsible for this asymmetry would be an influence of rotation on the observing height. Rotation causes a blue- or redshift of the spectral line, but the filter positions of the filtergraph of HMI are fixed. This might cause an asymmetry of the observing height in the east-west direction. It would be helpful to have HMI data with a different roll angle for studying these asymmetries.

Such an asymmetry of the shrinking-Sun effect is also present in the synthetic data derived from the STAGGER simulations. Unfortunately, the flow velocities derived from the STAGGER data have a high noise level due to the short length of the time-series, so it is not clear whether this observed asymmetry is just caused by noise or whether it is a real signal. A real asymmetry would be hard to explain, since the STAGGER simulations do not exhibit any large asymmetries. There is only a weak large-scale flow and the magnetic field in the photosphere has no preferred horizontal direction.

It might be possible to remove the shrinking-Sun effect from measurements of large-scale flows by observing at different $B_0$-angles. In CCD coordinates,  solar rotation and meridional flow depend on the $B_0$-angle (in a well known manner), while the shrinking-Sun effect should be constant. Unfortunately, the distance of SDO to the Sun is not constant in time. This directly affects the spatial resolution of the HMI data (in km). It also causes variations of the temperature of the front window of HMI which should lead to variations of the PSF. We expect these effects to cause a time-variation of the shrinking-Sun effect. Such an effect would be significantly worse for {\it Solar Orbiter}, due to the highly eccentric orbit.

The origin of the shrinking-Sun effect and other systematic errors can be studied with the {\it Solar Orbiter} mission. Combined observations of the same region on the Sun from two different vantage points, e.g., with {\it Solar Orbiter} and another instrument, could be used to derive an empirical correction for the shrinking-Sun effect.

\begin{acknowledgements}
We acknowledge support from Deutsche Forschungsgemeinschaft (DFG) through SFB 963/1 "Astrophysical Flow Instabilities and Turbulence" (Project A1). Support was also provided by European Union FP7 projects SPACEINN and SOLARNET. The German Data Center for SDO, funded by the German Aerospace Center (DLR), provided the IT infrastructure for this project. L.G. acknowledges support from the Center for Space Science, NYU Abu Dhabi Institute, UAE.
\end{acknowledgements}

\bibliographystyle{aa} 
\bibliography{literature} 

\end{document}